# On-chip beam rotators, adiabatic mode converters, and waveplates through low-loss waveguides with variable cross-sections

Bangshan Sun[1*], Fyodor Morozko[2], Patrick S. Salter[1], Simon Moser[3], Zhikai Pong[1], Raj B. Patel[4,5], Ian A. Walmsley[4], Mohan Wang[1], Adir Hazan[2], Nicolas Barré[3], Alexander Jesacher[3,6], Julian Fells[1], Chao He[1], Aviad Katiyi[2], Zhen-Nan Tian[7], Alina Karabchevsky[2*] and Martin J. Booth[1,6*]

[1]Department of Engineering Science, University of Oxford, Oxford OX1 3PJ, United Kingdom
[2]School of Electrical and Computer Engineering, Ben-Gurion University of the Negev, P.O.B. 653, Beer-Sheva, 8410501, Israel
[3]Institute of Biomedical Physics, Medical University of Innsbruck, Müllerstraße 44, 6020 Innsbruck, Austria
[4]Ultrafast Quantum Optics group, Department of Physics, Imperial College London, London, United Kingdom
[5]Department of Physics, University of Oxford, Oxford, United Kingdom
[6]Erlangen Graduate School in Advanced Optical Technologies (SAOT), Friedrich-Alexander-University Erlangen-Nürnberg, Paul-Gordan-Straße 6, 91052 Erlangen, Germany
[7]State Key Laboratory of Integrated Optoelectronics, College of Electronic Science and Engineering, Jilin University, Changchun 130012, China

*Corresponding to: Bangshan Sun (bangshan.sun@eng.ox.ac.uk), or Alina Karabchevsky (alinak@bgu.ac.il), or Martin J. Booth (martin.booth@eng.ox.ac.uk))

## Abstract

Photonics integrated circuitry would benefit considerably from the ability to arbitrarily control waveguide cross-sections with high precision and low loss, in order to provide more degrees of freedom in manipulating propagating light. Here, we report a new method for femtosecond laser writing of optical-fibre-compatible glass waveguides, namely spherical phase induced multi-core waveguide (SPIM-WG), which addresses this challenging task with three dimensional on-chip light control. Fabricating in the heating regime with high scanning speed, precise deformation of cross-sections is still achievable along the waveguide, with shapes and sizes finely controllable of high resolution in both horizontal and vertical transversal directions. We observed that these waveguides have high refractive index contrast of 0.017, low propagation loss of 0.14 dB/cm, and very low coupling loss of 0.19 dB coupled from a single mode fibre. SPIM-WG devices were easily fabricated that were able to perform on-chip beam rotation through varying angles, or manipulate polarization state of propagating light for target wavelengths. We also demonstrated SPIM-WG mode converters that provide arbitrary adiabatic mode conversion with high efficiency between symmetric and asymmetric non-uniform modes; examples include circular, elliptical modes and asymmetric modes from ppKTP (periodically-poled potassium titanyl phosphate) waveguides which are generally applied in frequency conversion and quantum light sources. Created inside optical glass, these waveguides and devices have the capability to operate across ultra-broad bands from visible to infrared wavelengths. The compatibility with optical fibre also paves the way toward packaged photonic integrated circuitry, which usually needs input and output fibre connections.

## Introduction

Low-loss integrated optical waveguides have indispensable roles in wide range of significant modern technologies, such as integrated photonic chips[1–3], quantum applications[4–9], and topological photonics[10–14]. In recent years, these research areas have underpinned some of the most remarkable advances in modern applied physics. The waveguides produced with silica-on-silicon technology (silica waveguides on a silicon chip, or SoS) hold significant future promise for both high-performance photonic chips[2] and also quantum computing[4,6–9]. Thousands of integrated components can be fabricated on a single wafer using a similar technological process as employed in industrial micro-electronic chips, with availability to robustly control the phase of single elements rendering the chip re-programmable. In a complementary manner, femtosecond laser (fs-laser) waveguide writing[15,16] has found a unique position for producing rapid prototypes and three-dimensional (3D) waveguide arrays. Laser written waveguides have been demonstrated to have significant potential in areas of such as topological photonics[10–14,17], as well as quantum technology applications[18–23]. Compared to other technologies, femtosecond laser material processing holds powerful advantages in creating complicated structures[24], demonstrating wide applications in various interesting areas, including opto-fluidic chips[25], photonic crystals[26], micro-ablation[27], nanolithography[28] and many others.

For most current applications, based simply on light guiding or cross-coupling, the waveguide transection remains fixed along the propagation direction, primarily because it is very difficult to find an efficient fabrication method that has the ability to arbitrarily transform cross-section shape and size along a waveguide with high resolution while maintaining low loss. Silica-on-silicon waveguides are normally fabricated through etching on a two-dimensional (2D) plane, where it is non-trivial to transform the transection in 3D. Precise cross-section deformation is also an open challenge for low loss fs-laser written waveguides in glass. Existing reported flexible cross-section transformation in glass substrate waveguides employed the powerful "classic multi-scan" technique[29–31], where fine control is possible in one transversal direction with resolution of ~0.4 µm, but accompanied by a typical coarse resolution, for an example, 8 µm in the other transversal direction. The coarse resolution lies along the direction of the fs-laser propagation, where the extended focal volume creates a complicated refractive index (RI) structure in the glass with both positive and negative regions of index change, making it difficult to build precisely controlled RI structures along this direction from a simple stack of multi-scans when they are spatially too close with each other. In comparison, a 3D stack of closely spaced multi-scans is much easier in other materials where the fs-laser helps to create a much simpler positive refractive index region[32]. For example, in recent innovative works demonstrating waveguides by two-photon polymerisation of SU-8 resin, various complicated cross-section structures[33] could be produced and transformed along beam propagation direction[34]. An exceptional high polarization conversion efficiency of >90% was achieved by twisting the shape of SU-8 resin waveguides. This excellent work by Prof. Sun's group[34] also demonstrated high polarization efficiency can be achieved by twisting waveguide in both visible region and infrared region.

Many applications in integrated optics would benefit from waveguides whose transection could be arbitrarily transformed in order to have much more versatile light control. The polarization state of light carries important information in most photonic chip-based applications, such as polarization encoding for quantum information[35], polarization division multiplexing in optical communications[36], and polarization sensors[37]. Though implementing polarization manipulation in free space is easy, arbitrary on-chip polarization control using integrated waveguides remains an open challenge. Existing on-chip polarization rotation technologies are mostly based on etching and deposition[38–44], with high performance devices operating with a bandwidth up to 100 - 200 nm. Created on a planar layer, the devices sometimes require high design precision and accuracy in fabrication. On the other hand, by using femtosecond laser writing, it was reported that the waveguide's birefringence could be finely tuned in fused silica[45,46]. Osellame's group reported excellent work with polarization conversion in glass to create rotated waveplates in optical waveguides[47]. By controllably rotating the birefringence axis, the waveguides were demonstrated for quantum state tomography.

Mode conversion is another well-known concept in integrated photonics, typically used to transform the size of the guided mode to suit different parts of the device via tapered waveguides[48–51]. For efficient mode conversion, it is desired that the transition is gradual such that the process is adiabatic. Existing mode converters are mostly based on similar etching and deposition technique, seeing limitations in operating bandwidth, spatial dimension and fabrication complexity. There have been reports in femtosecond laser written waveguides, where the power of the fabrication laser was continuously tuned to vary the size of the guided mode[52,53]. Conversion between different symmetric guided LP0x modes is also possible through more complex structuring[54]. However, there is still lack of flexible mode shape and precise size control, and there is yet to be a solution to address mode conversion between the asymmetric guided modes with high precision.

Considering these limitations, it is beneficial to find a universal solution to address these outstanding challenges in waveguide-based integrated photonic circuits. In this paper, we firstly demonstrate the concept of spherical phase induced multi-core waveguides (SPIM-WGs), which enables glass waveguides with high refractive index contrast and high precision in arbitrary 3D cross-section shape control. We then characterize loss properties with varying waveguide parameters and demonstrate arbitrary on-chip beam rotation. Following these, we present SPIM-WGs' capability in manipulating the polarization state of light. Finally, we demonstrate high efficiency arbitrary adiabatic mode conversions including asymmetric modes together with advanced mode matching.

## Results

### Spherical Phase Induced Multi-Core Waveguides (SPIM-WGs)

Historically, fs-laser written waveguides operate with two different fabrication regimes: non-heating regime with low laser repetition rate (< 10 kHz); and heating regime with high laser repetition rate (> 500 kHz). These two fabrication regimes involve different processes to create the RI profile in glass[15,55] (Supplementary Note 1). We develop our technology based on the latter, which has much higher fabrication efficiency, reducing processing time from several hours to 1-2 minutes. Higher refractive index contrast (refractive index difference between waveguide core and substrate) can also be obtained which helps in our applications.

In the rest of this paper, we refer to a common Cartesian coordinate system (Fig. 1 (a) and inset of Fig. 2 (a)) in which the waveguide was written along the *y* direction. The *z* axis (vertical direction) points along the optical axis of the laser-writing system. The *x* axis (horizontal direction) is perpendicular to the *y* and *z* axes. As briefly mentioned above, the "classic multi-scan" technique[29–31] is a powerful method for creating rectangular shape waveguides in glass. The fabrication laser focus is scanned multiple times through the glass substrate with a *x* axis separation of about 0.4 μm between each pass to build up the waveguide RI profile. As shown in Fig. 1 (a), in the vertical *z* axis dimension, control over the cross-section is challenging since the *z* axis separation is limited (to ~8 μm, depending on objective NA). To enable fine control of sizes, and ultimately achieve arbitrary deformation of cross-section, it is essential to find a solution to reduce the step size along the *z* direction, without affecting the uniformity of overall refractive index profile.

One might test the idea, illustrated in Fig. 1 (b)-Scheme I, where all laser spot scans closely stack together both along *x* and along *z*. However, as shown by the LED-illuminated transmission microscope image in Scheme I, the fabricated feature includes two large dark areas surrounded by bright regions, indicating a complicated structure of both positive and negative refractive index, which is most likely due to local thermal accumulation during processing[15,55]. During heating regime fs-laser fabrication, the high repetition rate laser creates a localized region of high intensity plasma. If the multiple scans are chosen to be too close with each other (e.g., 0.4 μm in both the *x* and *z* directions), the high intensity plasma generated by a newer scan melts the nearby feature created by existing scans, producing complicated structures. As shown in Fig. 1 (b)-Scheme I, when the waveguide is tested with laser transmission, a non-uniform multi-mode profile was observed.

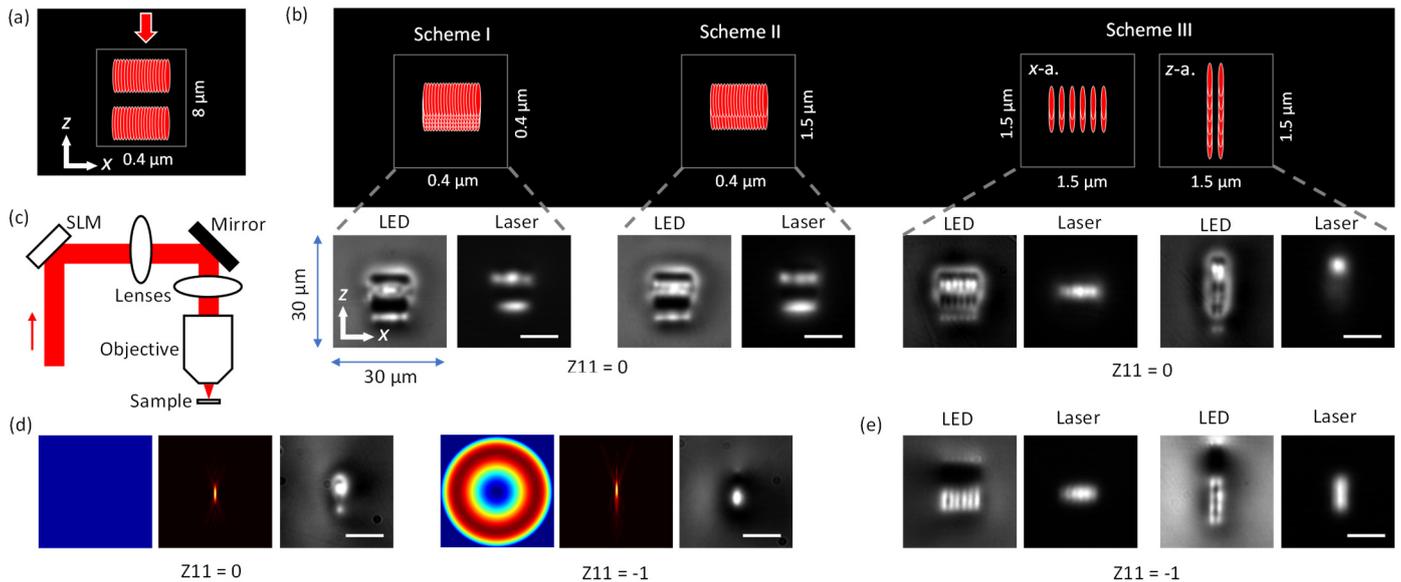

Figure 1. **Enabling fine control of waveguide cross-section shape and sizes by SPIM-WGs.** All images (LED, laser, simulation) excluding the phase pattern have the same frame size of 30×30 μm. Scale bars are 10 μm. (a) Classic multi-scan laser fabrication technique has a large $z$ step resolution (8 μm in this case). The red arrow marks laser propagation direction. Distances of core spacing along $x$ axis ($\Delta x$), and $z$ axis ($\Delta z$) are presented beside the diagram. (b) Three proposed fabrication schemes and the corresponding fabricated waveguides. In experimental demonstrations, Scheme I had $\Delta x = \Delta z = 0.4$ μm, with 20 horizontal and 6 vertical scans. Scheme II had $\Delta x = 0.4$ μm, $\Delta z = 1.5$ μm, with 20 horizontal and 2 vertical scans. Scheme III had $\Delta x = \Delta z = 1.5$ μm, with 6 horizontal and 2 vertical scans for the $x$-aligned rectangular waveguide, 2 horizontal and 6 vertical scans for the $z$-aligned rectangular waveguide. Labels: "LED" – images obtained with LED illuminated microscope; "Laser"- 785 nm laser transmission mode profile imaged at the waveguide output facet; "Z11" – manually induced primary spherical aberration[56] (corresponding to Zernike mode 11 phase aberration) applied to spatial light modulator (SLM). "x-a." - x-aligned rectangular waveguides. "z-a." - z-aligned rectangular waveguides. (c) Simplified diagram of laser fabrication system with phase control. SLM is imaged to objective pupil with a 4f telescope system. SLM: spatial light modulator; Objective: objective lens. (d) Manually induced additional spherical beam shaping phases to SLM and their effects on laser focus and single scan waveguides. Note that these phases are not used to correct spherical aberration caused by refractive index mismatch between air and sample, which was pre-corrected in all our experiments (please refer to "Methods- Phase pattern for SLM" section for details). Left: beam shaping phase applied to SLM in addition to aberration corrections. Middle: simulated focal intensity distribution (enlarged images in Supplementary Fig. S1). Right: LED illumined microscopic image of single scan waveguides. (e) Scheme III fabrication with negative spherical beam shaping phase (Z11 = -1), demonstrating significant improvement of cross-section control.

Observed from a single laser scan in Fig. 1 (d), we noticed that the refractive index asymmetry only exists along $z$ direction but not $x$ direction (which is consistent with existing reports[53,57,58]). We therefore explored the fabrication in Scheme II to see whether increased $z$ scanning step of 1.5 μm from 0.4 μm could help. While there was a slight improvement, as shown in Fig. 1 (b)-Scheme II, where positive RI region become wider along $x$ direction, the transmitted laser mode was still far away from design. We therefore relaxed both horizontal and vertical scanning steps to 1.5 μm in Scheme III, and fabricated both $x$-aligned rectangular and $z$-aligned rectangular waveguides. As shown by the images in Fig. 1 (b), Scheme III gave us better processed features. Comparing the laser mode profiles of all the three schemes in Fig. 1 (b), waveguides fabricated by Scheme III presents a confined one-lobe laser guiding region instead of two laser guiding regions shown in Scheme I and II. We explored varying fabrication parameters (pulse energy 50-150 nJ, scanning speed 1-12 mm/s) and cross-section sizes, however neither Scheme I nor Scheme II was able to fabricate a feature close to the design.

For all the waveguide fabrications in this paper, system induced aberrations were corrected by a wavefront sensorless adaptive optics method using a liquid crystal spatial light modulator (SLM) integrated into the laser fabrication system[59,60]. We experimentally verified the correction by the imaging of laser focus before each fabrication session. The spherical aberration that arose from refractive index mismatch between immersion and sample, was also pre-corrected by using the SLM[61]. It was important to ensure these aberrations were well corrected before our exploration (more details in Method section). For the convenience, we expressed this situation as "Zernike mode 11 equals 0" or "Z11 = 0", where the 11th Zernike polynomial mode corresponds to the lowest order spherical aberration.

While Scheme III looked to give best results for fabrication strategy (Fig. 1 (b)), there were however still large areas of complicated refractive index structure around the waveguide core. For the *z-aligned* rectangular waveguide, only the top region could guide light (bottom right images in Fig. 1 (b)). Again, the problem was due to the fact that asymmetric complicated refractive index structure of single-scan waveguide was along the *z* direction[58]. To resolve this problem, we introduced additional wavefront shaping, allowing us to gain a powerful capability to simplify and precisely control the fs-laser modified refractive index structure. The primary spherical aberration Zernike mode[56] was chosen, as it can change laser focal shape along the *z* axis, while maintaining circular symmetry in the *x-y* plane (more analysis in supplementary Fig. S1). We found that by deliberately introducing a negative spherical aberration phase, we were able to relocate more energy to the bottom half of the laser focus, shifting heat distribution along the *z* direction, thus producing a considerably simplified refractive index structure (Fig. 1 (d), Supplementary Fig. S1). Through extensive investigation, we found there was a difference in the waveguide formation process, that an applied spherical phase appears to limit heat accumulation at the top of laser focus, making it possible to generate a modified refractive index structure that matches the shape of laser focus. In comparison, using a conventional single scan without spherical aberrations, the shape of the modified refractive index structure was normally different to the laser focus. It was found empirically that an amplitude of -0.8 to -1.3 radians rms for Z11 worked well, so we chose negative one radian rms (Z11 = -1) for subsequent fabrication. As shown in Fig. 1 (d), fabrication with Z11 = -1 greatly simplified the single scan generated refractive index structure. Applying fabrication with Z11 = -1 to multi-scan *x-aligned* and *z-aligned* rectangular waveguides of Scheme III, we were able to produce structures which were well-matched to the original waveguide design (Fig. 1 (e), more results in Supplementary Fig. S2 and Fig. S3). The guided mode profiles showed well-confined elliptical modes.

Using manually induced spherical phase, we fabricated waveguides with core distances down to 0.3 μm and saw negligible difference in loss and mode properties. The distance of core spacing should be chosen based on practical applications. Larger spacing greatly increases fabrication efficiency and can be more suitable for longer wavelength applications. Smaller spacing gives higher resolution in control over cross-section shape and size and may be used for shorter wavelengths. As we targeted the design of devices to be optimized at the wavelength of 1550 nm, 1.5 μm was chosen for the core distance of most waveguides in this paper. In Supplementary Fig. S2, we summarized and compared the fabrication of waveguides with varying core spacing from 0.5 μm to 3.5 μm. SPIM-WGs technique created waveguides with much better cross-section than classic multiscan technique for all the core spacings, and most importantly it enabled new capability to create waveguides with core spacing <2 μm, which is typically not easy through the classic multiscan method. Besides, the minimal sizes of SPIM-WGs created by 0.5NA objective lens with 514 nm laser are around 3×0.5 μm. These minimal sizes are determined by the NA of objective lens and the wavelength of fabrication laser. It is possible to create SPIM-WGs with cross-section sizes smaller than 2×0.3 μm with a >0.7NA objective lens.

We named the waveguides fabricated by this technique as spherical phase induced multi-core waveguides (SPIM-WGs). We experimentally compared an alternative solution - to simply adopt a high NA objective lens in "classic multi-scan" (Fig. 1 (a)), which could reduce the vertical separation down to ~3 μm with a 1.3NA oil objective lens. However, our experimental results suggested that SPIM-WGs still hold several competitive advantages, including much higher fabrication efficiency (an order of magnitude reduction in fabrication time), more control over refractive index profile, and lower propagation loss. Some details of these will be analysed further in this paper.

**Beam Rotators Through Twisted Waveguides**

Using the capability to finely control cross-section size along all transversal axes, we created waveguides with cross-section shape and size that varied along the length. The concept of these waveguides is illustrated in the schematics of Fig. 2 (a), where a single waveguide is composed of a straight *z-aligned* rectangular waveguide region at the input, twisted waveguide region at the middle, and *x-aligned* rectangular waveguide region at the output. To realize the concept in experiment, waveguides were fabricated with 9×2 multiple scans, with each scan translating continuously through the entire sample. The transition in the twisted region from *z-aligned* rectangular to *x-aligned* rectangular was achieved by a smooth rotation of the 9×2 array by 90 degrees along a length of 1.4 mm. A transmission microscope image viewing from top of the fabricated sample (2D projection) shows the clear change in transverse waveguide dimension in the twisted region.

To demonstrate the laser mode transition efficiency, we tested the twisted waveguide with both 785 nm and 1550 nm lasers. As shown in Fig. 2 (b), when laser light was coupled from a *x-aligned* rectangular facet (right side of waveguide), the mode profiles obtained at the output of waveguide were rotated 90 degrees to become *z-aligned*. Similarly, when light was coupled from the *z-aligned* rectangular facet, the output mode profiles were rotated 90 degrees to be *x-aligned* (more results in Supplementary Fig. S3).

To accurately characterize the waveguide refractive index profile at various points along the device, measurements were made using 3D tomographic microscopy[62]. From the profiles in Fig. 2 (c), we can see that the multi-core waveguides along the *z* axis were well combined and formed a smooth transition of positive refractive index regions along *z* (the vertical bright lines). The positive refractive index in each single core was highly uniform, significantly reducing scattering to achieve low waveguide loss. The shape and size of the positive refractive index regions were highly consistent across the whole multi-core cross-section. The light guiding region is highlighted with a dashed box; and there are surrounding areas of negative refractive index, which could further enhance mode confinement. As one of the SPIM-WGs' advantages, the refractive index contrast was measured to be $17\times10^{-3}$, which is remarkably higher than that of most reported glass waveguides[15]. We believe this benefitted from both spherical phase control and partial overwriting (more details in Supplementary Fig. S4). Compared to a classic multi-scan technique, the refractive index distribution of SPIM-WGs is better organized and highly predictable, contributing to much better waveguide qualities, especially low losses, which will be discussed later.

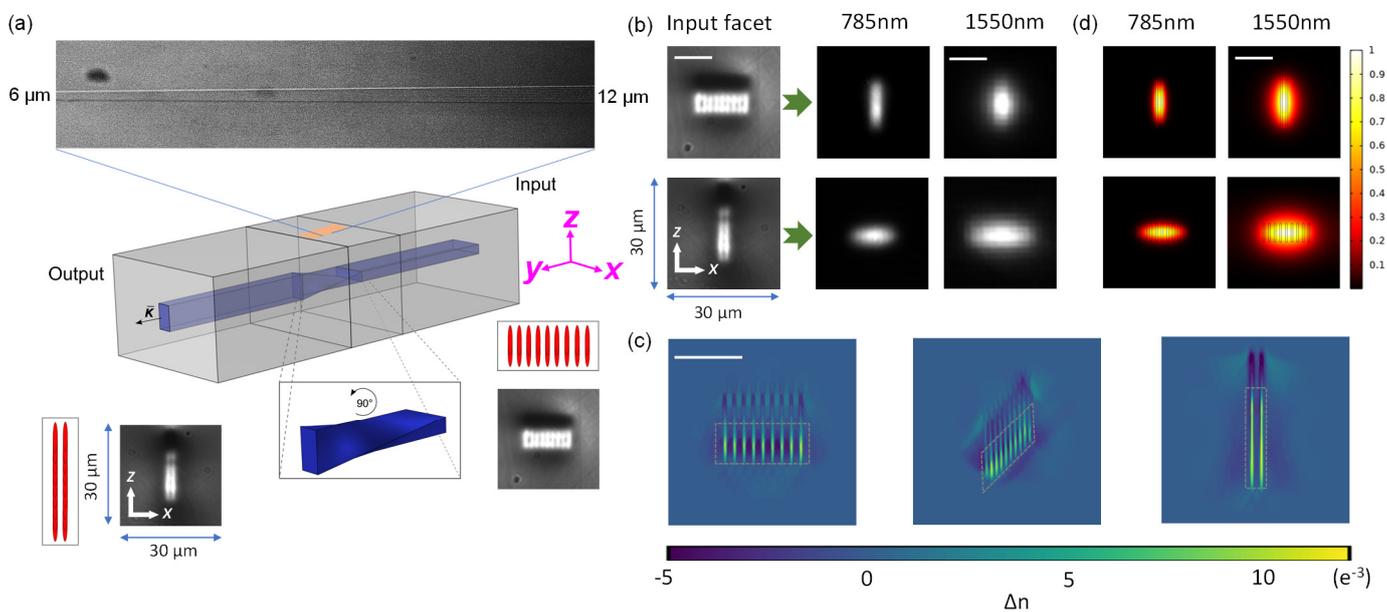

Figure 2. **Characterization of twisted shape SPIM-WGs.** All images (LED, laser, simulation) have the same frame size of 30×30 μm. Scale bars are 10 μm. (a) Composition of a twisted waveguide. The lengths of *z-aligned* rectangular, twisted, *x-aligned* rectangular regions were 9.3 mm, 1.4 mm, 9.3 mm, respectively for the studied waveguide. LED microscopic image top view of the fabricated sample is included, where a transition from 6 μm width to 12 μm width is clearly seen from the top 3D to 2D projection. Multi-scans of the laser spot at waveguide facets are shown as schematics. Coordinates: *x/z* transversal axis, *y* longitudinal axis. (b) Rectangular beam rotation can be achieved by a twisted waveguide, with 785 nm laser and 1550 nm laser tested respectively. Top: light guided from *x-aligned* rectangular input facet was converted to *z-aligned* rectangular modes. Bottom: light guided from *z-aligned* rectangular input facet was converted to *x-aligned* rectangular modes. (c) Measured refractive index profiles by a 3D tomographic microscope. Dashed boxes highlight positive refractive index regions which was able to guide laser light. Refractive index contrast was measured as high as 0.017. (d) COMSOL simulations of mode intensity distribution for 785 nm laser and 1550 nm laser. The simulations were conducted based on measured refractive index data in the positive region from tomographic microscope (highlighted by dashed boxed).

In order to investigate the waveguiding properties from spatially separated regions of positive index change, we conducted simulations using the experimentally measured refractive index data. As shown in Fig. 2 (d), the guiding modes at both wavelengths were uniform without any evidence of deterioration. We note that the mode profile was dependent on spacing of multi-core as well as refractive index contrast. In our design, choosing 1.5 µm core distance with $17\times10^{-3}$ refractive index contrast was sufficient to produce a uniform laser guiding mode with excellent confinement for both the 785 nm and the 1550 nm laser. As seen from Fig. 2 (b) and (d), the mode size was smaller for the shorter wavelength, which is expected. Based on a large number of simulations for varying waveguide cross-sections, we found by appropriate design that these periodic positive-negative refractive index transection profiles were well able to produce mode profiles similar to a homogeneous step index waveguide, and had negligible impact to waveguide losses, which will be discussed in the following text.

**Light guiding performance**

It is essential to evaluate waveguide losses to inspect whether the multi-core structure or the twisted region may introduce additional losses to the waveguide. In order to have a comprehensive understanding of losses, we fabricated several sets of waveguides with different cross-section shapes and sizes, as illustrated in Fig. 3 (a). We here used waveguides fabricated with a single pass of laser focus[15] (referred to here as "classic single-scan waveguide") as the basis for comparison. Loss measurements were conducted by a cut-back method (details in Methods section) at 785 nm wavelength. We first evaluated whether the multi-core structures introduced additional loss compared to the classic single-scan waveguide. The average measured waveguide propagation losses are summarized in Fig. 3 (b). The numbers were averaged across several waveguides with the same configuration to show repeatability. Except for the *x-aligned* rectangular waveguide with 10×4 µm cross-section, all the other SPIM-WGs had propagation losses lower than that of classic single-scan waveguide. We therefore concluded that overall, SPIM-WGs had lower propagation losses than that of classic single-scan waveguides. We believe this is due to higher refractive index uniformity and contrast of SPIM-WGs.

We then investigated whether the twisted region inside SPIM-WGs introduced additional losses compared to the straight region. In the measured results of Fig. 3 (b), we found twisted waveguides have on average a propagation loss that lies between that of *x-aligned* rectangular waveguides and *z-aligned* rectangular waveguides. It was therefore concluded that additional loss induced by waveguide cross-section twisting should be negligible. On the other hand, twisting SPIM-WGs were demonstrated with even lower propagation losses than classic single-scan waveguides in Fig. 3 (b). We note that for single scan waveguides, the conventional method with $Z_{11} = 0$ provides slightly lower loss and better circularity, while SPIM-WGs with $Z_{11} = -1$ hold advantages in nearly all the aspects whenever a multiscan approach is needed.

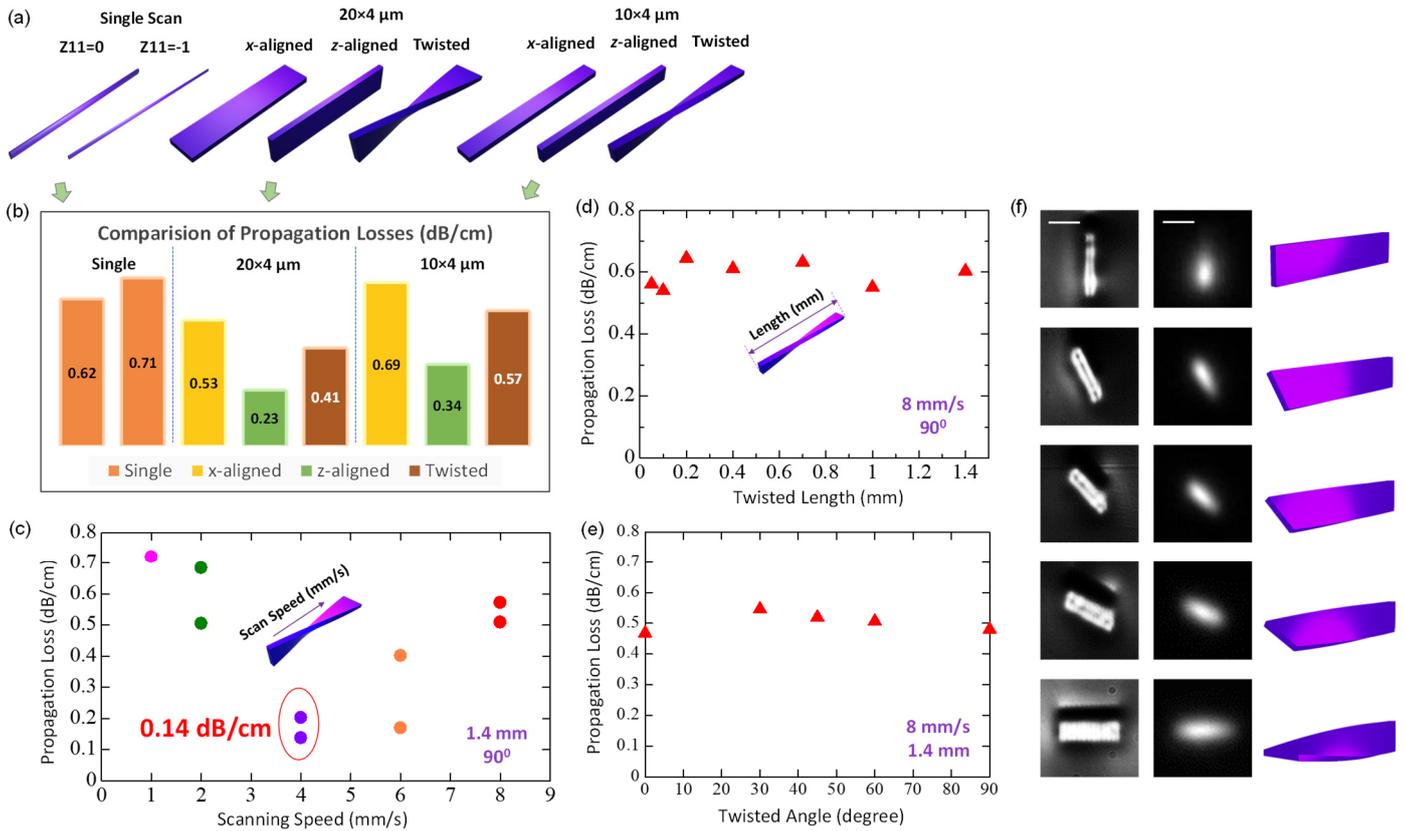

Figure 3. **Light guiding performance of SPIM-WGs.** All images (LED, laser) have the same frame size of 30×30 μm. Scale bars are 10 μm. (a) Sets of waveguides were fabricated with different cross-section shapes and sizes. Classic single-scan, *x-aligned*, *z-aligned* waveguides were two waveguides per set, while twisted waveguides (twisted length of 1.4 mm) were four (two *z-aligned* input facet, two *x-aligned* input facet) per set. Cross-section images are included in Supplementary Fig. S3. (b) Comparison of propagation losses for classic single-scan, *x-aligned*, *z-aligned* and twisted waveguides. Each number presented in the figure is an average of 2 measured waveguides for single-scan, *x-aligned*, and *z-aligned* sets; an average of 4 waveguides for twisted sets. The fabrication speed was 8mm/s. (c) Twisted waveguides' propagation losses versus laser focal spot scanning speed. Cross-section size was 20×4 μm, twisted length was 1.4 mm, twisted angle was 90°. (d) Twisted waveguides propagation losses versus twisted region length. Cross-section size was 10×4 μm, laser scanning speed was 8 mm/s (sacrificing loss to gain fabrication efficiency), twisted angle was 90°. (e) Twisted waveguide propagation losses versus twisted angle. Cross-section size was 20×4 μm, laser scanning speed was 8 mm/s (sacrificing loss to reduce fabrication time), twisted length was 1.4 mm. (f) Demonstration of beam rotation by several sets of twisted waveguides with varying twisted angle. Left: LED transmission microscopic images of waveguide facet. Middle: measured 1550 nm laser mode profile images. Right: diagrams of twisted SPIM-WGs with different angles.

We conducted experiments to evaluate how the fabrication parameters affected the performance of twisted waveguides. Fig. 3 (c) summarized how the twisted waveguide's propagation loss could be optimized by changing laser scanning speed. It appeared that with pulse energy of 78nJ, a scanning speed around 4 mm/s was optimal. A waveguide fabricated with these parameters was measured with propagation loss as low as 0.14 dB/cm, which is close to the limit of EAGLE glass absorption at 785 nm[63]. During our experiments, we constantly measured propagation losses in the range of 0.13-0.2 dB/cm for twisted shape SPIM-WGs fabricated with 4 mm/s scanning speed, further confirming these low propagation losses were easily repeatable.

We fabricated waveguides with different twisted region lengths to evaluate how this parameter affects the overall performance. As shown in Fig. 3 (d), the total propagation losses (straight plus twisted regions) remained almost constant when reducing the length of twisted region, which means the SPIM-WGs with twisted region as short as 0.05 mm had similar overall loss to one with a 1.4 mm twisted region. Moreover, we found that the waveguides with different twisted lengths had comparable performance in *z-aligned* to *x-aligned* mode conversion. It is notable that waveguides in Fig. 3 (d) were fabricated with higher scanning speed of 8 mm/s, which reduced fabrication time by half. Optimizing scanning speed could easily bring down the losses as indicated in Fig. 3 (c).

Finally, we demonstrated SPIM-WGs' flexibility in controlling the twisted angles, providing versatile beam rotation capability. We fabricated twisted waveguides starting from a *z-aligned* rectangular shape into twisted angles of 0°, 30°, 45°, 60°, 90°. Fig. 3 (f) includes both LED illuminated microscopic images and 1550 nm laser transmission mode profiles, showing not only good control over cross-section shape, but also the flexibility to rotate the orientation of elliptical laser guiding modes. Moreover, as demonstrated in Fig. 3 (e), varying the twisted angle did not have noticeable effects on SPIM-WGs' propagation losses.

**Adiabatic Mode Converters with Advanced Mode Matching**

We demonstrate that SPIM-WGs enable a new capability to flexibly create mode converters that can arbitrarily transform modes regardless of their symmetry. There are many photonic chip applications that require mode manipulation, for example, when mode matching is needed. We highlight four common application cases to demonstrate this capability. In most optical chips, coupling of laser light in/out of a single mode fibre is important[3]. In terms of direct laser written waveguides, researchers have adopted methods to achieve higher coupling efficiency[52,53] by controlling the fabrication laser power. Based on SPIM-WGs, we created converters whose cross-sections were transformed between circular and rectangular shapes (diagram on the left of Fig. 4 (a)). The mode conversion performance is demonstrated in Fig. 4 (a), where the mode intensity plots indicate a clear transition between circular and elliptical modes. Both the circular or rectangular shapes and sizes can be flexibly and precisely controlled. In our experiments, the circular facet of the mode converter was designed to have the same physical dimensions as the core of the 1550 nm single mode fibre (a diameter of 8 μm). This provided excellent mode matching; we observed that 95.7% light was coupled from the fibre tip to the circular facet of SPIM-WGs mode converter. This represents a significant decrease of coupling loss after advanced mode matching, from 1.69 dB in the case of rectangular facet to 0.19 dB in the case of circular facet based on our measurement with 1550 nm laser. The total loss (coupling + propagation) of a mode converter with length of 1.32 cm was measured to be remarkably as low as 0.59 dB, where the waveguide was fabricated with scanning speed of 6 mm/s. During our experiments, the total losses of 10 converters fabricated in the same configuration were measured to be in the range of 0.59 - 0.75 dB, confirming high consistency of light guiding performance.

As a second example, we consider ppKTP (periodically-poled potassium titanyl phosphate) waveguides[64–68], which are used in nonlinear optics, in particular for frequency conversion and quantum light sources. The waveguiding mode is defined via a rubidium ion-exchange process beginning at the surface of the material and penetrating below with gradually reduced concentration. This typically leads to a skewed-Gaussian mode profile[69–71], inducing high coupling losses to single-mode fibre due to mode mismatch, which is a major hurdle for effective integration of these devices. Based on our previous measurements, the coupling loss between single-mode fibre and a ppKTP waveguide is around 70%[67]. We observed that improving efficiency from 70% to >80% for each mode would allow one to beat the shot-noise limit in phase-sensing without post-selection[67]. We demonstrate that our SPIM-WGs technique can easily create a converter to transform between a ppKTP waveguide mode and a circular mode from an external single mode fibre, therefore provide a significantly improved coupling. We designed a refractive index profile, presented in Fig. 4 (b), that was able to generate a mode profile matched to the mode from a typical ppKTP waveguide. The designed refractive index profile was created by fine tuning the size and shape with feedback from COMSOL simulation. We fabricated this type of mode converter from an 8 μm diameter circular shape (matching to the core size of single mode fibre) to the designed ppKTP refractive index with sizes of 8×8 μm, 6×6 μm, and 4×4 μm. A typical mode converter with length of 1.38 cm was measured to have total loss remarkably as low as 0.65 dB, with coupling loss of 0.24 dB (94.6% light coupled). In total, 18 converters were fabricated, and their total losses were measured in the range of 0.65-0.8 dB. The mode profiles of one converter with mock ppKTP waveguide of 8×8 μm size are presented in Fig. 4 (c), where the mode shape transition is clearly seen. In practical applications, mode converters could be cut to be shorter (e.g., ~0.5 cm), easily reducing the propagation loss and raising overall efficiency to be > 90%.

The advantage of creating adiabatic mode converters inside glass is that devices can be designed for a particular wavelength across a broad range from visible to near infrared, as glass has low absorption in these wavelengths. To demonstrate the applicability of our technique in another wavelength band, we created another two mode converters (presented in Fig. 4 (c) and (d)) for 785nm laser conversion. Mode converters for other target wavelengths would also be easily achievable, as the design would differ mostly in size.

As the third example, we created a mode converter that can convert between a Gaussian mode and a rectangular TE10 mode. We excited the TE01 mode with two lobes along *x* direction, however it is easy to switch to the design of the other orientation (*z*-aligned) as we demonstrated above (sometimes TE01 and TE10 are used to distinguish the orientation of the lobes). This conventional TE01 mode has the same mode intensity profile as LP11 modes in waveguide theory[72], meaning TE01 mode we generated can be coupled to either a rectangular waveguide or a circular waveguide. Either a TE01 mode (Fig. 4 (c)), or an elliptical shape mode (e.g., Fig. 2) can be excited at the rectangular shape output facet. The transition between them can be easily controlled by either a slight shift in the input angle (~5°) or a slight shift in the position (~0.5-1 µm) of input beam relative to the waveguide. In practical applications, we used optical glue to fix the angle or position between fibre and waveguide sample. As the demonstration in Fig. 4 (c), we chose to shift the angle, which also brought slightly higher propagation loss. The total loss (coupling + propagation) of this mode converter with length of 1.27 cm was measured to be 0.78 dB at 785nm laser, where the waveguide was fabricated with scanning speed of 8 mm/s.

As the fourth example, we created a mode converter that can convert between circular shape Gaussian mode and a circular TE10 mode, which has a ring shape intensity with a hollow core[73,74]. In order to generate a ring shape intensity with symmetric and uniform intensity, we used COMSOL simulations to precisely design the position for each single laser scan (in total 18 scans for one converter). We note the position of each single scan is not evenly distributed along the circle, since the structure generated by single scan is elongated along the *z* direction. This is seen in the last COMSOL simulated image of Fig. 4 (f). As the result, the measured ring shape mode profile at 785nm has high symmetry and uniform intensity. The total loss (coupling + propagation) of the mode converter with length of 1.27 cm was measured to be 0.73 dB at 785nm laser, where the waveguide was fabricated with scanning speed of 8 mm/s. In total, we created 24 mode converters with ring diameter varying from 6.5 µm to 13 µm, with total insertion loss measured in the range of 0.73 to 0.88dB.

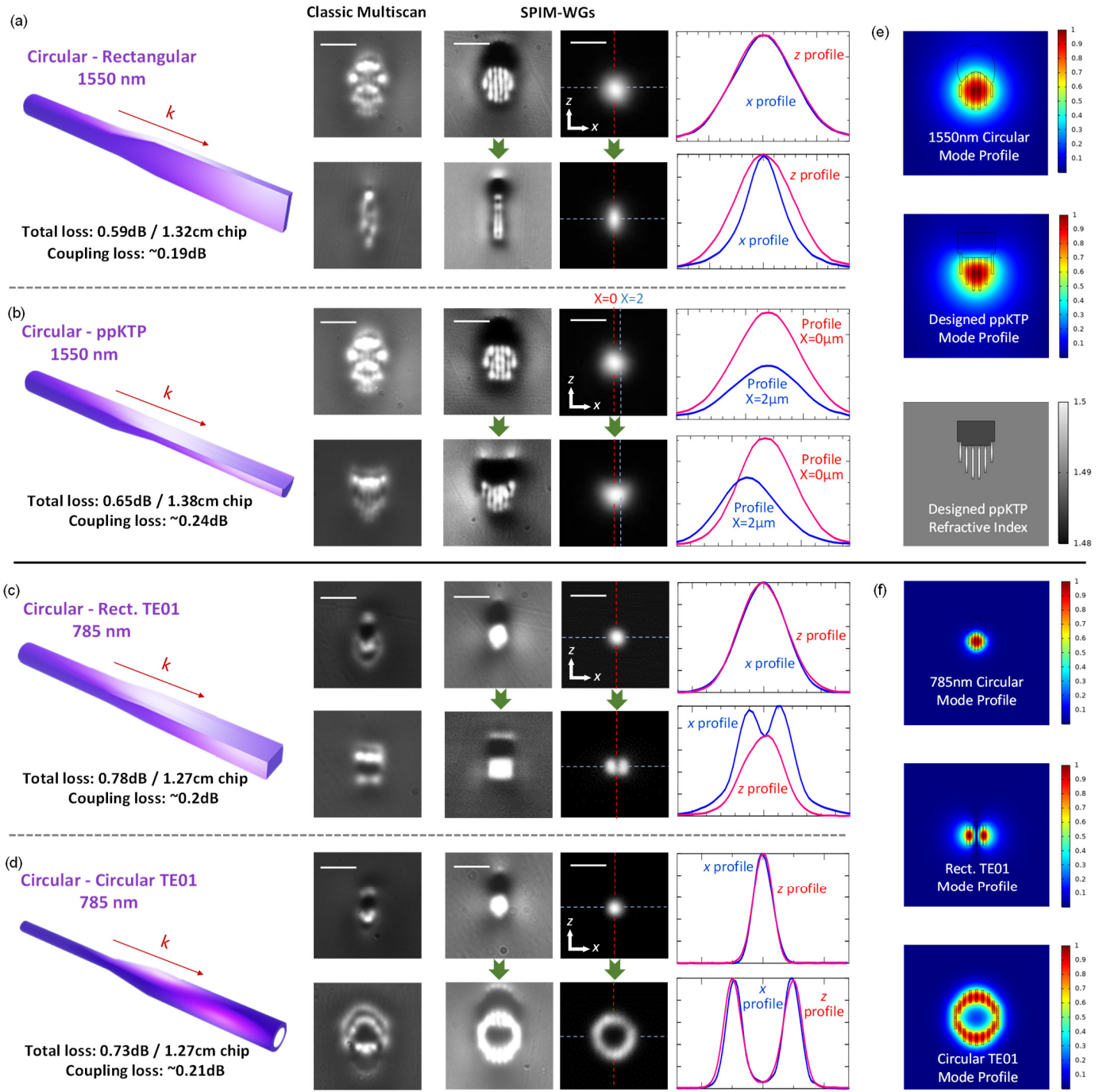

Figure 4. **Adiabatic mode converters with advanced matching of arbitrary modes**. All images (LED, laser) in (a) and (b) have the same frame size of 30×30 μm. Scale bars are 10 μm. (a) Left: Circular-Rectangular mode converter that couples single mode fibres (circular shape mode) to rectangular shape waveguides (elliptical shape mode). A waveguide with length of 1.32 cm was measured to have ultra-low total loss of 0.59 dB (coupling + propagation). The coupling loss was calculated to be 0.19 dB (95.7% light coupled). Middle: LED illuminated microscopic images demonstrating classic multiscan technique is not capable to fabricate these high-quality mode converters while SPIM-WGs technique enables this capability. 1550 nm laser mode profiles are presented for SPIM-WGs. Right: plots of intensity versus distance for corresponding SPIM-WGs circular and elliptical mode profiles. Plotting lines are indicated in mode profiles. (b) Circular-ppKTP mode converter that couples single mode fibres to ppKTP waveguides. A waveguide with length of 1.38 cm was measured at 1550nm wavelength to have total loss of 0.65 dB (coupling + propagation). The coupling loss was 0.24 dB (94.6% light coupled). (c) Circular-Rectangular TE01 mode converter that converts between Gaussian circular mode with rectangular TE01 mode (also called LP11 mode in waveguide theory). A waveguide with length of 1.27 cm was measured at 785nm wavelength to have ultra-low total loss of 0.78 dB (coupling + propagation). The coupling loss was calculated to be as low as 0.2 dB (95.5% light coupled). (d) Circular-Circular TE01 mode converter that converts between Gaussian circular mode with circular TE01 mode or a hollow ring shape intensity distribution. A waveguide with length of 1.27 cm was measured at 785nm wavelength to have total loss of 0.73 dB (coupling + propagation). The coupling loss was calculated to be 0.21 dB (95.3% light coupled). (e) Top: COMSOL simulated field distribution of designed circular shape mode profile to match a single mode fibre at 1550nm. Middle: COMSOL simulated field distribution of designed mode profile to match ppKTP waveguide at 1550nm. Bottom: designed refractive index profile for matching ppKTP waveguide mode. (f) Top: COMSOL simulated field distribution of designed circular shape mode profile to match a single mode fibre at 785nm. Middle: COMSOL simulated field distribution of designed rectangular TE01 mode profile at 785nm. Bottom: COMSOL simulated field distribution of designed circular TE01 mode profile at 785nm.

To demonstrate SPIM-WGs' capability in creating high quality adiabatic mode converters, we also added the fabrication results from classic multiscan for comparison in Fig. 4. As we can see that the structures created by classic multiscan are very complicated. The situation becomes even worse for the application in shorter wavelengths (Fig. 4 (c)(d)). Mode converters require precise fabrication, so that we can conclude that the classic multiscan fabrication in the heating regime is not capable of creating these mode converters.

The capability of SPIM-WGs in creating adiabatic mode converters is not restricted to the above four examples. We also found that the adiabatic transitions of the cross-sections introduced negligible additional loss in SPIM-WGs mode converters. The propagation losses of the mode converters were found to be nearly the same as the waveguides with fixed cross-section with same sizes. We also conducted experimental verification of adiabatic process for the mode converters, with results summarized in Supplementary Fig. S7.

**On-chip Wavelength Dependent Waveplates**

In this section, we extend the capability of SPIM-WGs by demonstrating the polarization state manipulation of guided linear polarized light, to enhance the toolbox of SPIM-WGs photonic circuits. To prove this concept on a chip, we constructed polarization-controlled experimental apparatus (details in Methods section). The waveguide was tested with linearly polarized light from either a monochromatic or wide band supercontinuum laser source. SPIM-WGs were studied to evaluate their polarization conversion efficiency (PCE) for propagated light, which is defined as,

$$PCE_{TE \to TM} = \frac{P_{TM}}{P_{TE} + P_{TM}}$$

$$PCE_{TM \to TE} = \frac{P_{TE}}{P_{TE} + P_{TM}}$$

in which, $P_{TE(TM)}$ is the power in the TE (TM) polarization at the waveguide output facet.

We investigated 90° twisted waveguides with varying twisted length and total length (diagram in Fig. 2 (a)). When tested with 0°/90° linear polarized 1550 nm monochromatic laser, we found that appropriately designed twisted waveguides were able to rotate the polarization of transmitted laser light, through the mechanism of adiabatic mode evolution in the twisted region[34,39,42,75]. When light propagates along the twisted region with such a rotated rectangular shape, not only does the mode shape undergo an adiabatic evolution, but the polarization of the photons also gradually changes together with the evolution of fundamental modes. Fig. 5 (a) compared two twisted waveguides with same total length of 30 mm, but different twisted length of 25 mm (top) and 15 mm (bottom). As seen in the mode images, the twisted waveguide with 25 mm twisted length maintained the same polarization state of guided laser light at 1550 nm. In comparison, the twisted waveguide with 15 mm length was able to convert the polarization state with efficiency up to 60% as demonstrated in bottom images of Fig. 5 (a). With a TE input, TM mode was strongly observed at the output of twisted waveguide, while a similar case was seen for TM mode conversion into TE mode.

The capability of polarization manipulation of twisted waveguides is demonstrated for near-infrared light in the results of Fig. 5 (b) and (c). We observed that the polarization conversion had periodic variation across a wide range of wavelengths. We think the oscillations can be the consequence of excitation and interference of higher order modes in our waveguides. Supplementary Note 2 provides details of the mathematical representation of this mode interference. With the definition of modal group index[76] $n_g = -\frac{\lambda^2}{2\pi}\frac{d\beta}{d\lambda} = n_{eff} - \lambda\frac{dn_{eff}}{d\lambda}$, where $\beta$ is the propagation constant, $\lambda$ is the wavelength of transmitted laser light and $n_{eff}$ is effective refractive index, we obtained the following expression for the period of spectral oscillations in wavelength (more details in Supplementary Note 2):

$$\Delta\lambda = \pm\frac{\lambda^2}{(n_{g1} - n_{g2})L}$$

$n_g$ has relatively weaker dependence over $\lambda$, so that $\Delta\lambda$ is almost a function of $\lambda^2$ and $L$. The obtained expression explains faster oscillations for waveguides with longer total length (larger $L$, comparing Fig. 5 (b) and (c)), and narrowing of the wavelength period at shorter wavelengths. The period of the observed oscillations may indicate that the interference occurs between modes of different order rather than between polarization modes of the same order. The energies of these modes are interfering and coupling along the length of the twisted region where both mode conversion and polarization conversion take place. Some mode interferences are constructive to polarization conversion, while others are destructive (details in Supplementary Note 2). In some particular wavelength bands, the destructive effects of mode interferences are heavier than that of other wavelength bands, where we observed nearly zero polarization conversion for these wavelengths. The fact that the polarization conversion is differently affected by mode interference explains the oscillation effect of PCE across wide wavelength ranges.

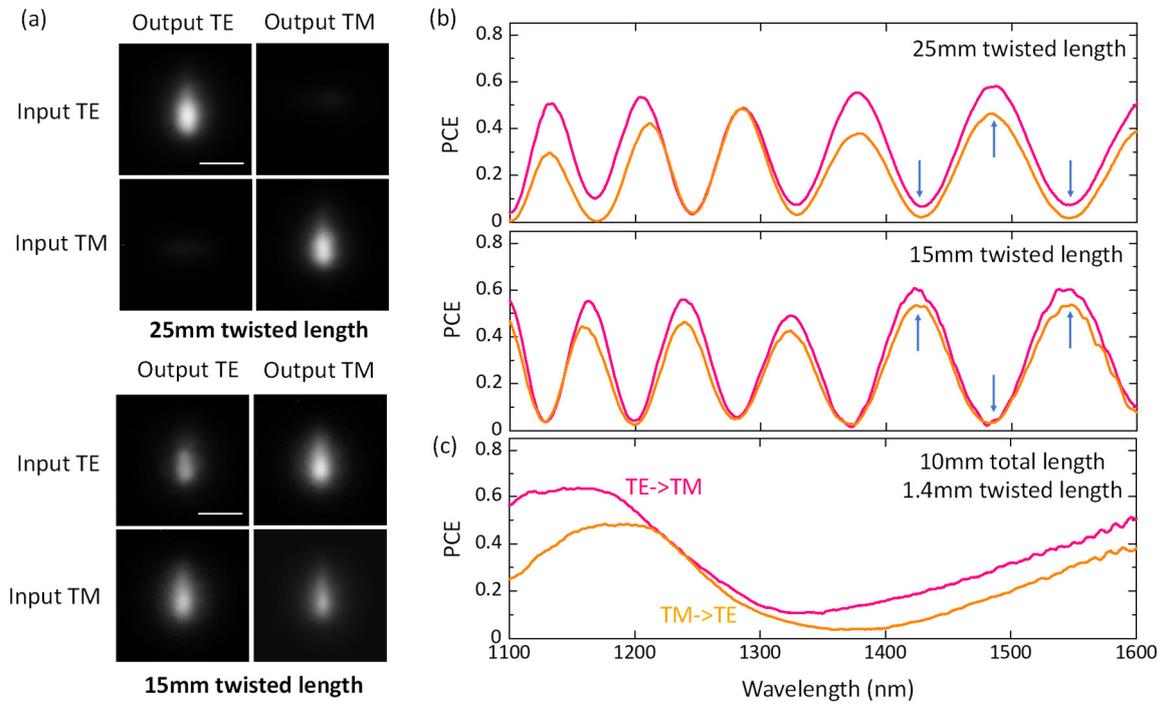

Figure 5. **Demonstration of on-chip polarization manipulation of SPIM-WGs.** All images (LED, laser) have the same frame size of 30×30 μm. Scale bars are 10 μm. (a) Results of polarization-controlled experiments at 1550 nm wavelength laser. Top: twisted waveguide with 25 mm twisted length. Bottom: twisted waveguide with 15 mm twisted length. Images are waveguide mode profiles observed for particular output polarization state, with input laser light of either pure TE or pure TM polarization. Both waveguides have total length of 30 mm. (b) Measured polarization conversion PCE for twisted waveguides with total length of 30 mm, different twisted lengths of 25 mm and 15 mm, at wide near-infrared wavelengths. PCEs are controllable for any target wavelength by modifying twisted length while fixing total length. Red curve is measured with horizontal polarized input laser, while orange curve is vertical polarized input laser. (c) Measured polarization conversion PCE for twisted waveguides with total length of 10 mm, twisted lengths of 1.4 mm, at wide near-infrared wavelengths. The polarization manipulation behaviour of straight rectangular waveguides was measured for comparison in Supplementary Fig. S9. In Fig. 5, Waveguide cross-section sizes are 20×4 μm for all. Twisted waveguides are with 90 degrees twisting angle.

Modification of the total waveguide length makes it possible to adjust the period of polarization conversion oscillation, which provides a useful tool for on-chip polarization manipulation. When the total waveguide length was fixed, but the length of the twisted region was changed, we found that the periodical oscillation can shift, as shown in Fig. 5 (b). Fixing the total length and controlling the length of twisted region thus becomes a second method for tunable polarization manipulation. Based on these observations, it is possible to design a waveplate which provides rotated polarization state at several target wavelengths while more or less maintaining original polarization state at the other wavelengths. For an example, as shown as middle figure of Fig. 5 (b), one waveplate with 15 mm twisted length operates as a waveplate at 1550 nm (telecom. C band) and 1420 nm (telecom. E band), but maintains original polarization state at 1490 nm (telecom. S band). Conversely, the waveguide with 25 mm twisted length (top figure) maintains the original polarization state at 1550 nm (telecom. C band) and 1420 nm (telecom. E band), but operates as a waveplate at 1490 nm (telecom. S band). It is thus possible to design and achieve a waveplate with desired rotated polarization angle for particular wavelength. We achieved a PCE of around 70% through further reduction of the spacing between multiple cores to 0.65 µm to produce a more uniform refractive index cross-section. As the first extensive report of experimental investigations on polarization conversion effect for the weak refractive index contrast 90-degree twisted waveguides embedded inside glass, these results show the polarization conversion effect seems to be weaker than those waveguides with a high contrast step index[34] if same twisted length is applied.

Though it is not easy to access a full polarization conversion, with low losses and multi-wavelength applicability, the designed waveplate could still be applied to some important applications which do not need full polarization conversion, such as, various integrated quantum entangled photons sources[23], quantum study of polarization photons[47] and many other photonic devices which need integration of an on-chip waveplate. Another capability arises with the wavelength dependence, where the waveguide acts as a rotated waveplate at one or several wavelengths and while barely effecting the polarization at other different wavelengths which might be interesting for frequency-division multiplexing (FDM) telecommunication systems. Thirdly, the waveplate can be designed as a polarization-maintaining beam rotator, where the beam intensity profile can be rotated 90° with little change of the original polarization state. These cases are experimentally shown in Fig. 5 (b) and (c) where the polarization conversion approaches zero but beam profile itself was rotated 90°. Finally, the functionalities of polarization conversion can be incorporated together with adiabatic mode conversion in a single waveguide design, for devices exhibiting multiple functionalities, which is important for the future development of ultra-compact integrated photonic circuits.

## Discussion

We have made several breakthroughs by introducing this kind of SPIM-WGs. Firstly, we open the capability to create waveguides in glass not only with precisely organized refractive index and low propagation loss, but also have arbitrarily variable cross-section in any shape and size (to the limits of writing laser spot), with high resolution precision control along both horizontal and vertical axis, and in the heating regime with high fabrication speed. Secondly, with advanced mode matching capability of SPIM-WGs, we achieved very high coupling efficiency (95.7%) from an external single mode fibre to an on-chip device. The optical fibre compatibility of SPIM-WGs could be an important solution for future packaged photonic integrated circuitry. Thirdly, this new kind of waveguide enabled the easy creation of adiabatic mode converters that can handle complicated asymmetric modes, for which significant difficulties remain using existing technologies. SPIM-WGs also enabled new capabilities where thermal regime classic multi-scan is struggle to do (Supplementary Fig. S11). Together with the broadband applicability from visible to near-infrared, these breakthroughs constitute a significant development in integrated waveguide technology that will open new interesting applications in on-chip photonics, quantum technologies and a lot of other related areas.

## Materials and methods

### Femto-second laser fabrication system

The laser waveguide fabrication system used a regenerative amplified Yb:KGW laser (Light Conversion Pharos SP-06-1000-pp) with 1MHz repetition rate, 514 nm wavelength, 170 fs pulse duration. A Spatial Light Modulator (SLM, Hamamatsu Photonics X10468) was aligned and imaged by a 4-f lens system to the pupil of objective lens. The power of the laser beam was controlled with a motorized rotating half waveplate together with a Polarization Beam Splitter (PBS). The laser beam at objective lens focus was circularly polarized. The glass sample was fixed on a three-axis air bearing stage (Aerotech ABL10100L/ABL10100L/ANT95-3-V) to control the movement to inscribe waveguides.

The results presented in this paper were fabricated in borosilicate glass (Corning EAGLE 2000) samples, as we found high repetition laser fabrication in heating regime works better in this glass compound. SPIM-WGs fabrications in fused silica glass have slightly greater complexity, likely due to the difference in material properties between these two glass compounds[77].

In this paper, if not specified, the waveguide fabrication parameters were: 0.5NA objective lens (~93% transmission), waveguide depth of 120 µm from surface of glass sample, scanning speed of 8mm/s, pulse energy (measured at the objective pupil) was 87nJ for fabrication with Z11 = 0, and 78nJ for fabrication with Z11 = -1. For all the fabrications in this paper, the primary spherical aberration, arising from the refractive index mismatch between immersion and sample, was corrected[61].

### Phase pattern for SLM

For all the waveguide fabrications in this paper, the phase pattern applied to SLM was,

$$\begin{aligned} SLM\ &Phase\ Pattern \\ &= Compensation\ of\ system\ induced\ aberrations \\ &+ Compensation\ of\ refractive\ index\ mismatch\ aberration \\ &+ Manual\ induced\ Z11\ beam\ shaping\ phase\ for\ SPIM\_WGs \end{aligned}$$

where "Z11" represents the first order Zernike polynomial mode for spherical aberration. For the convenience, we expressed the situation, that first two aberration terms (system induced + RI mismatch) were pre-corrected, as "Zernike mode 11 equals 0" or "Z11 = 0".

### Waveguide images and loss characterizations

After direct laser fabrication, waveguide samples were polished by using a sequence of 30 µm, 9 µm and 3 µm polishing films. A layer of at least 200 µm glass were polished off both input and output facets of the glass. Supplementary Fig. S5 (a) includes a system diagram for microscopic images and loss measurements. A self-built LED-illuminated widefield transmission microscope (microscope 1) was used to check waveguide cross-sections after polishing. To image laser guiding mode profiles, single mode fibres were used to guide 785 nm or 1550 nm laser light into the input facet of a waveguide. The fibre output was mounted and adjusted in a six-axis stage (three spatial axes plus three angle adjustment). Another self-built microscope (microscope 2) was used to monitor and measure the distance between fibre tip and waveguide input facet. During the measurement of coupling losses, we brought the fibre tip to be close (< 1µm) to the waveguide sample to avoid input beam expansion. The output facet of a waveguide was imaged by microscope 1 in order to capture guided laser mode profile.

To analyse the propagation loss and coupling loss of the waveguides, a cut-back approach was adopted. A power meter was placed at camera position of microscope to measure laser powers: 1) direct at fibre tip without waveguide sample ($P_{fiber}$); 2) at waveguide output facet after full-length waveguide ($P_{output\_long}$); 3) at waveguide output facet after cutting the waveguide to a shorter length ($P_{output\_short}$), where the lengths of waveguide ($Length_{long}$, $Length_{short}$) were measured in centimetres. The waveguide coupling efficiency ($E_c$) is defined as proportion of laser light coupled from the waveguide to the other waveguide/fibre. The propagation efficiency per centimetre ($E_p$) is defined as proportion of laser light transmitted after one centimetre of waveguide. $E_c$ and $E_p$ can be resolved by solving below two equations with the measurements of longer and shorter waveguides,

$$P_{output\_long} = P_{fiber} \cdot E_c \cdot (E_p)^{Length_{long}}$$

$$P_{output\_short} = P_{fiber} \cdot E_c \cdot (E_p)^{Length_{short}}$$

Applying the relation between Efficiency (*E*) and Loss (in terms of dB), $Loss = 10 \cdot \log_{10} E$, the coupling loss ($Loss_c$) and propagation loss ($Loss_p$) can also be directly calculated by resolving below,

$$P_{output\_long} = P_{fiber} \cdot 10^{Loss_c/10} \cdot 10^{Length_{long} \cdot Loss_p/10}$$

$$P_{output\_short} = P_{fiber} \cdot 10^{Loss_c/10} \cdot 10^{Length_{short} \cdot Loss_p/10}$$

We would like to note that a negative loss value in dB means power attenuation, while positive loss value represents power amplification. However, in waveguide optics, people normally report absolute loss values to represent waveguide losses. In this paper, we adopt this same waveguide convention, so that one must simply revert them to negative numbers in all their own calculations (for example, 0.59dB waveguide loss should be reverted to -0.59dB mathematically).

**Waveguide polarization measurements**

Supplementary Fig. S5 (b) includes the detailed diagram of the polarisation measurement system. To demonstrate polarization manipulation, the coupling-in was performed via the focusing objective. The light was collected from the output facet via objective and split into two paths by 50:50 beam splitter. Concurrently to the imaged-on camera (VIS-SWIR Ninox-640) mode profiles of twisted and reference-straight waveguides, the spectrum was collected by optical spectrum analyser (OSA Yokogawa 6370D). Two polarizers were placed before and after the waveguide sample in order to conduct the polarization conversion measurement. The two polarizers were placed with relative angle of 0º to characterize the polarization maintaining, and with relative angle of 90º to characterize the polarization conversion.

**Simulations**

A Fourier optics model was adopted to simulate (in Matlab) the focal intensity distribution of an objective lens focussing into glass[78–80]. The focal intensity distribution was calculated by solving the Rayleigh-Sommerfeld diffraction integral from the pupil illumination. The spherical phase was added into the model by a phase function φ in the pupil of the objective lens (Fig. 1 (d), Supplementary Fig. S1). To simulate waveguide mode profiles, we used finite element method (COMSOL Multiphysics) to create complicated refractive index structures which approximated the measured refractive index profiles from 3D tomographic microscope. When multiple modes existed, we presented the major mode.

**Refractive index profile measurement**

Refractive index profiles were measured by a self-built 3D tomographic microscope[62]. The imaging system records many intensity images of a waveguide at different illumination angles ranging between about -45° and 45° The light source was a collimated blue LED (460 nm). A two-dimensional refractive index cross section of the waveguide was then reconstructed from the image stack using an error reduction algorithm based on gradient descent and simulated beam propagation.

**Acknowledgements**

B.S. acknowledges valuable discussions with Prof. Robert R. Thomson in Heriot-Watt University, Dr. Josh Nunn in University of Bath, Prof. Dong Wu in University of Science and Technology of China, as well as valuable suggestions from all the anonymous reviewers. All their comments greatly helped to improve the paper.

This project was partially supported by the European Research Council Advanced Grants AdOMiS (695140), UK Engineering and Physical Sciences Research Council grants EP/T001062/1, EP/R004803/01, EP/T00326X/1, Austrian Science Fund (FWF) I3984-N36. A.K. acknowledges Israel Innovation Authority KAMIN #69073 'Development of mode converters technology with twisted waveguides on a chip'. C.H. acknowledges Junior Research Fellowship of St John's College in Oxford.

**Author contributions**

B.S conceived the idea of SPIM-WGs, developed the technique, and oversaw the project. B.S. designed and fabricated all the waveguides and devices for beam rotators, adiabatic mode converters and waveplates, characterized laser guiding performance, obtained microscopic images and conducted loss analysis. A.K., F.M., A.H., A.Kat. conceived the concept of twisted shape waveguides for waveplates, conducted analytical description, modelling, measurements of twisted waveguide's polarization conversion properties. S.M., N.B. and A.J. developed the tomographic microscopy, measured and analysed refractive index profiles. I.W., R.P., P.S., M.B. and B.S. conceived and developed the concepts of adiabatic mode converters. B.S. and Z.P. conducted COMSOL simulations. B.S. performed laser focus simulations. B.S. and A.H. drew the waveguide diagrams. M.W., J.F. contributed to various waveguide performance characterizations and discussions. C.H. contributed in polarization discussions and Z.T. contributed in waveguide and waveplate fabrication discussions. B.S. constructed all the figures and wrote the manuscript, M.B. and P.S. co-wrote the paper, with great help from A.K., R.P., F.M., and valuable comments from A.J., I.W. All authors discussed the results and reviewed the manuscript.

**Conflict of Interest**

The authors declare no competing interests.